\documentclass[review]{elsarticle}

\usepackage{amssymb}
\usepackage{latexsym}
\usepackage{url}
\usepackage{xcolor}
\usepackage{comment}
\usepackage{hyperref}

\usepackage{amsmath,mathtools}
\usepackage{array}
\usepackage{mdwmath}
\usepackage{booktabs}
\usepackage{algorithm}
\usepackage[noend]{algpseudocode}
\usepackage{amsfonts}
\usepackage{mdwtab}
\usepackage{eqparbox}
\usepackage{lineno,hyperref}
\usepackage{subfigure} 

\modulolinenumbers[5]
\usepackage{multirow}

\journal{Neurocomputing}

\begin{document}

\begin{frontmatter}

\title{CADA: Multi-scale Collaborative Adversarial Domain Adaptation for Unsupervised Optic Disc and Cup Segmentation}



\author[BME]{Peng Liu}
\author[ECE]{Charlie T. Tran}
\author[UNCC]{Bin Kong}
\author[BME,ECE,CAM]{Ruogu Fang\corref{mycorrespondingauthor}}
\cortext[mycorrespondingauthor]{Corresponding author}
\ead{ruogu.fang@bme.ufl.edu}

\address[BME]{J. Crayton   Pruitt   Family   Dept.   of   Biomedical Engineering, University of Florida, Gainesville, FL 32611, USA}
\address[ECE]{Dept.   of   Electrical and Computer Engineering, University of Florida, Gainesville, FL 32611, USA}
\address[CAM]{Center for Cognitive Aging and Memory, University of Florida, Gainesville, FL 32611, USA}
\address[UNCC]{Department of Computer Science, UNC Charlotte, Charlotte, NC 28223, USA}

\begin{abstract}
Recently, deep neural networks have demonstrated comparable and even better performance than board-certified ophthalmologists in well-annotated datasets. However, the diversity of retinal imaging devices poses a significant challenge: domain shift, which leads to performance degradation when applying the deep learning models trained on one domain to new testing domains. In this paper, we propose a multi-scale input along with multiple domain adaptors applied hierarchically in both feature and output spaces. The proposed training strategy and novel unsupervised domain adaptation framework, called Collaborative Adversarial Domain Adaptation (CADA), can effectively overcome the challenge. Multi-scale inputs can reduce the information loss due to the pooling layers used in the network for feature extraction, while our proposed CADA is an interactive paradigm that presents an exquisite collaborative adaptation through both adversarial learning and ensembling weights at different network layers. In particular, to produce a better prediction for the unlabeled target domain data, we simultaneously achieve domain invariance and model generalizability via adversarial learning at multi-scale outputs from different levels of network layers and maintaining an exponential moving average (EMA) of the historical weights during training. Without annotating any sample from the target domain, multiple adversarial losses in encoder and decoder layers guide the extraction of domain-invariant features to confuse the domain classifier. Meanwhile, the ensembling of weights via EMA reduces the uncertainty of adapting multiple discriminator learning. Comprehensive experimental results demonstrate that our CADA model incorporating multi-scale input training can overcome performance degradation and outperform state-of-the-art domain adaptation methods in segmenting retinal optic disc and cup from fundus images stemming from the REFUGE, Drishti-GS, and Rim-One-r3 datasets.
 \textit{The code will be available at \url{https://github.com/cswin/CADA}}.
\end{abstract}

\begin{keyword}
\texttt{Domain adaptation}\sep \texttt{Unsupervised learning} \sep \texttt{Segmentation} \sep \texttt{Retinal fundus images}

\end{keyword}

\end{frontmatter}


\section{Introduction}

Early diagnosis is vital for the treatment of various vision degradation diseases~\cite{regan1984low}, such as glaucoma, Diabetic Retinopathy (DR), and age-related macular degeneration. Many eye diseases can be revealed by the morphology of Optic Disc (OD) and Optic Cup (OC)~\cite{almazroa2015optic}. For instance, 
glaucoma is usually characterized by a large Cup to Disc Ratio (CDR), the ratio of the vertical diameter of the cup to the vertical diameter of the disc. Currently, determining CDR is mainly performed by pathology specialists. However, it is extremely expensive to accurately calculate CDR by human experts. Furthermore, manual delineation of these lesions also introduces subjectivity, intra-  and  inter-variability~\cite{danielewska2014glaucomatous}. Therefore, it is essential to automate the process of calculating CDR. OD and OC segmentation are commonly adopted to automatically calculate the CDR. Nevertheless, 
OD segmentation is challenging because pathological lesions usually occur on OD boundaries, which affect the accurate identification of the OD region. Accurate OC segmentation is more challenging due to the region overlap between the cup and the blood vessels~\cite{almazroa2016novel}.

\begin{figure}[h]
  \begin{center}
  \includegraphics[width=3.0in]{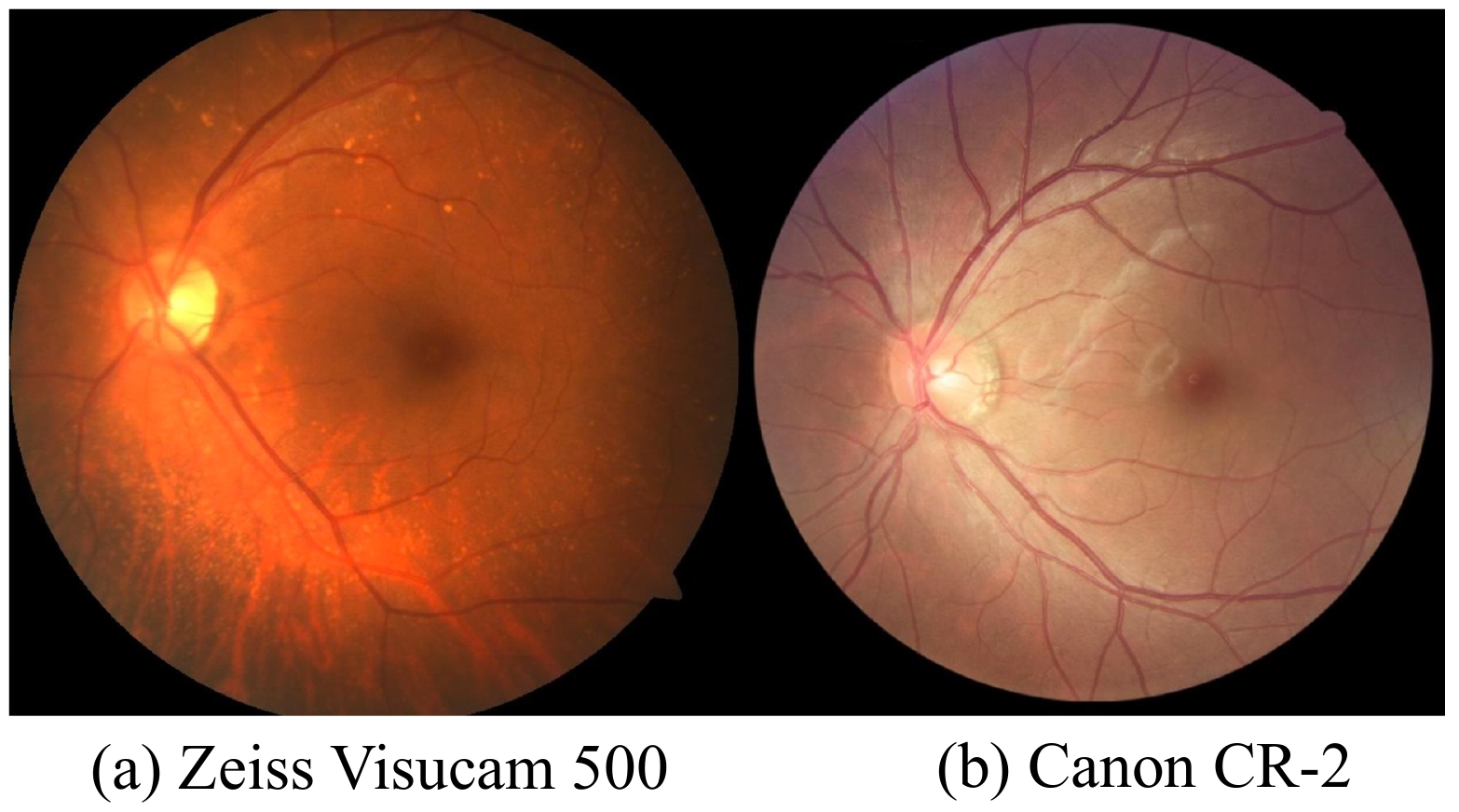}
  \caption{{Retinal fundus images collected by different fundus cameras revealing a variation in color and intensity.}}
  \label{fig: domains}
  \end{center}
\end{figure}

Recently, deep learning-based methods~\cite{fu2018joint, sevastopolsky2017optic, lim2015integrated, tan2017segmentation} have been proposed to overcome these challenges and some of them, e.g., M-Net~\cite{fu2018joint}, have demonstrated impressive results. Although these methods tend to perform well when being applied to well-annotated datasets, the segmentation performance of a trained network may degrade severely on datasets with different distributions, particularly for retinal fundus images captured with different imaging devices (e.g., different cameras, as illustrated in Fig.~\ref{fig: domains}). 
The variance among the diverse data domains limits deep learning's deployment in reality and impedes us from building a robust application for retinal fundus image parsing. To recover the degraded performance, annotating the fundus images captured from every new domain and then re-training or fine-tuning a model is an easy way but extremely expensive and even impractical for the medical areas that require expertise. 

To tackle this challenge, recent studies~\cite{ganin2014unsupervised,kamnitsas2017unsupervised,tsai2018learning,tzeng2017adversarial,liu2020pdam,liu2020unsupervised,wu2019vessel} have demonstrated the effectiveness of using deep learning for unsupervised domain adaptation to enhance the performance of applying models on unlabeled target domain data.  Existing works have mainly focused on minimizing the distance between the source and target domains to align the latent feature distributions from different domains~\cite{tzeng2017adversarial}. Several primary approaches can guide the alignment process, which includes image-to-image translation of the input images~\cite{hoffman2017cycada},  adversarial training for the intermediate representations in the layers of the model (encoder or decoder)~\cite{ganin2016domain}, and applying adversarial learning to the output of the model~\cite{tsai2018learning}. However, adversarial discriminative learning usually suffers from the instability of its training.  Numerous methods have been studied to tackle this challenge. Self-ensembling~\cite{laine2016temporal} is one of them recently applied to visual domain adaptation~\cite{french2017self}. In particular, gradient descent is used to train a student network, and the exponential moving average of the weights of the student is transferred to a teacher network after applying each training sample.  The mean square difference between the outputs of the student and the teacher is used as an unsupervised loss to train the student network. The paradigm of student-teacher has been a widely used strategy for unsupervised training of a deep neural network, feature extraction~\cite{he2020momentum}, and knowledge distillation~\cite{hinton2015distilling}. This unsupervised training strategy allows the student network to capture more information about the data during training and achieve a better prediction.

Furthermore, most of the existing methods have not considered providing multi-scale information of the data to the deep neural networks for having a better understanding of the difference between target and source domain features. In addition, most deep neural networks employ pooling layers to reduce model parameters and extract important features. However, pooling layers lead to a significant loss of the information in the original input data. In addition, the optic cup and disc in fundus image have high variance in brightness, color, shape, and orientation, which makes single-scale and single-level adversarial adaptation insufficient. Capturing only information from the output space neglects the intuition that low-level features are similar across various domains, leading to poor domain adaptability and missed opportunities for incorporating hierarchical features in segmentation predictions. To overcome these problems, in this study, we propose a multi-scale input training strategy to integrate different scales of features into different levels of the network layers. On one hand, multi-scale inputs performing at different levels of network layers can reduce the information loss due to the pooling layers in the network; On the other hand, it can provide rich information for the network to easily distinguish the difference between the source and target domain features. 

These multi-scale inputs are integrated into an encoder-decoder structure to form multiple sub-networks with multi-outputs. In this way, we can have multiple discriminators between the source and target domain input and leverage ensembling internally on the multiple distinctions at different levels of the network layers to have a better final segmentation prediction in target domain data. This proposed multi-domain adaptor approach can overcome the limitation of the current methods while comprehensively applying domain adaptation at hierarchical multi-scale in both feature and output space.

In this paper, we propose a novel unsupervised domain adaptation framework called Collaborative Adversarial Domain Adaptation (CADA) to further the state-of-the-art in overcoming the underlining domain shift problem. In particular, we take advantage of self-ensembling to stabilize the adversarial discriminative learning 
of the latent representations from domain shifting to prevent the network from getting stuck in degenerate solutions. 
Most importantly, we apply the unsupervised loss by adversarial learning not only to the output space but also to the input space and the intermediate representations of the network. 
Thus, from a complementary perspective, adversarial learning can consistently provide various model space and time-dependent weights to self-ensembling to accelerate the learning of the domain-invariant features and further enhance the stabilization of adversarial learning, forming a benign collaborative circulation and unified framework. 
The significant contributions of this paper are as follows: 
\begin{itemize}
    \item We propose CADA, a novel unsupervised domain adaptation framework, which exploits collaborative adversarial learning and weights self-ensembling for feature adaptation to tackle domain shift in a mutually beneficial and complementary manner at different network layers, resulting in a robust and accurate model. 
    \item We propose a multi-scale input training strategy to overcome the information loss when applying pooling layers in the network and offer an opportunity to integrate various scales of low-level and high-level features for improved network learning.
    \item We optimized feature adaptation by applying adversarial discriminative learning in two phases of the network, i.e., intermediate representation space and output space. More specifically, we apply adversarial learning at multiple layers of the network to learn the invariant features in both encoder and decoder phases simultaneously. 
    \item We reduce the uncertainty of multiple discriminators collaborative learning for domain shift via the EMA to ensemble model weights dynamically during training.
    \item We evaluate the effectiveness of our CADA on the challenging task of unsupervised joint segmentation of the retinal OD and OC. Our CADA can overcome performance degradation to domain shift and outperform one of the state-of-the-art domain adaptation methods with a significant performance gain on various datasets.
\end{itemize}
 
This work is a substantial extension of our conference paper ``CFEA: Collaborative Feature Ensembling Adaptation for Domain Adaptation in Unsupervised Optic Disc and Cup Segmentation''~\cite{liu2019cfea} published in Medical Image Computing and Computer Assisted Intervention (MICCAI) 2019.  In this extension, we substantially expanded our framework's reliability and scalability of overcoming domain shift issue in fundus images. In particular, we demonstrate the significant new contributions as below:
\begin{itemize}

\item We propose a novel multi-scale input layer to enhance the feature interaction between the encoder and the decoder where CFEA only uses a single scale. An input on each scale provides original image information to an encoder layer, which is followed by a decoder layer at the same network ``pyramid'' level. The rich original pixel-wise feature can infuse the interaction between encoder and decoder at the different feature-learning levels in the network. This infusion triggered by the multi-scale input can further guide the model learning and promote performance by reducing the significant information loss due to the pooling layers applied in the network and reducing the high variance of the optical cup and disc images in brightness, color, shape, and orientation.

\item Instead of a single domain adaptor (e.g., a discriminator network) at the end of the network in CFEA, we propose to apply multi-domain adaptors at hierarchical multi-scales in both feature and output space, which encourages the network to learn the domain-invariant features consistently. More importantly, they can collaboratively distinguish robust latent features in the scenarios that the optical cup and disc images have high variance in brightness, color, shape, and orientation, thus leading to a reliable and scalable domain adaptation framework. 

\item Comprehensive ablation studies are performed to investigate the effectiveness of the proposed framework. The ablation study investigates the importance of the encoder adversarial discriminative adaptation, the power of weights self-ensembling adaptation, the scalability of using multiple domain adaptors, and the choice of various combinations of the weights of loss functions. 

\item Evaluation on multiple public datasets is performed to show generalizability and stability of the proposed CADA framework compared to state-of-the-art methods.

\end{itemize}
\section{Related Work}

\subsection{Optic Disc and Optic Cup Features } 
The features of the Optic Disc (OD) and Optic Cup (OC) are critical for the diagnosis of eye diseases~\cite{almazroa2015optic}.  For example, ophthalmic pathologies (e.g., glaucoma) can be indicated by the variations of the shape, color, or depth of OD. The Cup to Disc Ratio (CDR), the ratio of the vertical diameter of the cup to the vertical diameter of the disc, is considered a valuable feature for diagnosing eye diseases~\cite{beck1987anterior}, such as glaucoma,  because higher CDR is highly associated with detectable visual field damage~\cite{le2003risk}. Currently, determining CDR is mainly performed by pathology specialists. However, it is expensive to calculate CDR by human experts accurately.
Furthermore, the variance of determining the CDR among professionals is usually significant, which can be caused by both the diversity of retinal fundus images and different experiences of the professionals~\cite{trucco2013validating}.  Therefore, it is essential to automate the process of calculating CDR.  On one hand, this automated process can reduce the cost of diagnosis. On the other hand, it can stabilize  diagnostic accuracy and improve the efficiency of retinopathy screening procedures.

\subsection{OD and OC Image Segmentation}

Image segmentation is a long-term research topic in the field of computer vision and image analysis. It is the basis for feature recognition and quantitative feature understanding~\cite{yuheng2017image}. 
In medical imaging, image segmentation is particularly important since it can help to locate related lesions/tumors and provide quantitatively analytical results of shapes/morphologies for clinicians. 
For example, image segmentation can automatically detect the OD and OC regions and calculate the CDR simultaneously, e.g.,~\cite{almazroa2016novel}. The task of OD segmentation is to detect the region between retinal and the rim. The presence of pathological lesions on the OD boundaries become problematic for OD detection. More to the point, OC detection is hindered by the region overlap between the cup and the blood vessels, as well as the color intensity changing between the cup and rim. It is critical to erase these challenges for reducing incorrect OD and OC segmentation that may cause a false diagnosis.

Recently, many deep learning-based studies~\cite{almazroa2015optic,almazroa2016novel, fu2018joint} have been proposed to overcome these challenges. In general, there are several steps to achieve a decent result. Firstly, a pre-trained disc center localization method~\cite{fu2018disc} is used to detect the OD and OC. The localization mainly acts as an attention mechanism so that the network can focus on essential regions and meanwhile, the polar transformation amplifies the relevant features to enable a more accessible learning process. Secondly, the localized areas are transformed (e.g., cropped, re-sized, and image coordinate system consistency) into the segmentation model training stage. Lastly, these transformed image regions are fed into an encoder-decoder convolutional network to predict the actual OD and OC regions for an arbitrary fundus image. The encoder is utilized to extract rich image features; the decoder part is used to produce accurate segmentation results based on the encoded features.  These combined techniques can reduce the negative effect on model performance caused by the variance in retinal images.  However, the variation is only constrained within one image domain, in which the training and testing images usually have similar distributions, such as background color and intensity. In practice, the testing images can be acquired from different types of cameras and have varying background or image intensity (as illustrated in Fig.~\ref{fig: domains}). The performance of a model trained on the dataset collected from one domain is severely degraded in another domain. This issue is referred to as ``domain shift''~\cite{wang2018deep}. It is critical to overcome this issue for a generalized and robust model in medical practice.

\subsection{Unsupervised Domain Adaptation} 
Saenk et al.~\cite{saenko2010adapting} originally introduced the unsupervised domain adaptation problem in tackling the performance degradation caused by the domain shift.  In particular,  unsupervised domain adaptation aims to tackle domain shift via adapting the training process of a model in an unsupervised manner. The model is adapted to improve the performance on the target domain. More importantly, leveraging unsupervised learning can reduce the tremendous and expensive data labeling work for the target domain.  Therefore, unsupervised domain adaptation is a promising direction to solve domain shift problems, especially, in the medical field whose data is multi-modal and requires expensive and expertise data labeling.

Recently, many deep learning-based domain adaptation methods~\cite{almazroa2016novel,ganin2016domain,taigman2016unsupervised} have been proposed and achieved several encouraging results. 
Many of these methods tackle the domain shift issue by extracting invariant features across the source and target domains. A critical approach for reducing the domain discrepancy is adversarial learning~\cite{ganin2014unsupervised}, which has become a fundamental method to obtain invariant information across multiple domains. In particular, it leverages the gradient discrepancy between learning the labeled and unlabeled data to minimize performance degradation. The implementation can either be image-to-image translation~\cite{zhu2017unpaired,isola2017image,liu2017unsupervised} in a convolutional neural network (CNN) input-end or multiple adversarial learning~\cite{,tsai2018learning, radford2015unsupervised} applied at the output-end of a CNN. Noticeably, the image-to-image translation usually introduces artifacts, which may be not a proper approach in the medical field. Therefore, in this work, we focus on gradient-based adversarial learning. 

\begin{figure}[t]
  \begin{center}
  \includegraphics[width=\textwidth]{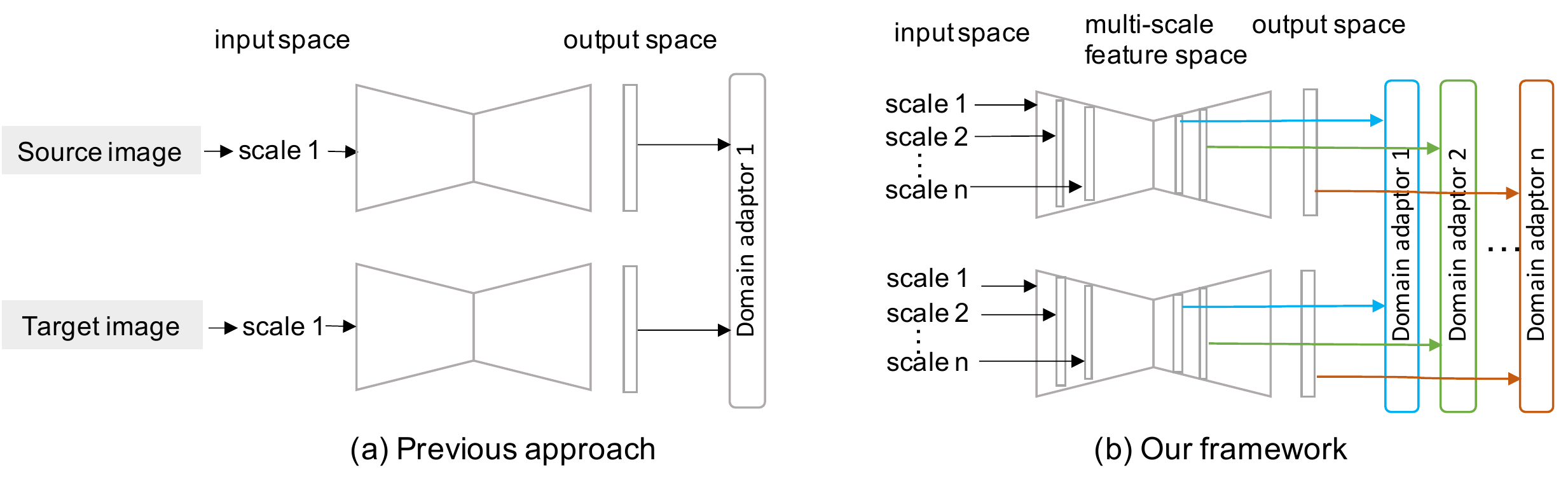}
  \caption{{Comparison between prior domain adaptation methods and our framework. Each colored arrow represents a connection with each multi-scale output to a corresponding domain adaptor. }}
  \label{fig:idea_diff}
  \end{center}
\end{figure}

In addition,  previous adversarial learning based approaches~\cite{wang2019patch} are mainly applied domain adaptation at the output space of a deep neural network. However, accurate image segmentation requires the model to capture both low- and high-level image representations; thus, it would be ideal for applying domain adaptation at various feature spaces with multi-scale representations. In this study, we comprehensively investigated whether employing multiple domain adaptors at a different level of layers of the network can benefit the representation learning across domains compared to the approaches only focusing on adaptation at the output space. We showed the design difference between previous methods and our framework and illustrated the major novelty in Fig.~\ref{fig:idea_diff}.

Furthermore, although adversarial learning can align the latent feature distribution of the source and target domain and have achieved encouraging results~\cite{ganin2014unsupervised}, the results of multiple adversarial learning-based methods are easily suffering from sub-optimal performance due to the difficulty of stabilizing the training process of multiple adversarial modules. Thus, in this work, we propose to leverage the Exponential Moving Average (EMA)~\cite{french2018self} computing method to dynamically ensemble learning weights as embedding multiple adversarial modules in a network. Meanwhile, this stabilization can bring not only a more robust but also an accurate model to effectively overcome the domain shift issue in fundus image segmentation problems.

\section{Multi-scale Collaborative Feature Ensembling Adaptation}
\subsection{Problem Formulation}
Unsupervised domain adaptation typically refers to the scenario: given a labeled source domain dataset with distribution $P(X_s)$ and the corresponding label $Y_s$ with distribution $P(Y_s|X_s)$, as well as a target dataset with distribution $P(X_t)$ and unknown label with distribution $P(Y_t|X_t)$, where $P(X_s)\neq P(X_t)$, the goal is to train a model from both labeled data $X_s$ and unlabeled data $X_t$, with which the expected model distribution $P(\hat{Y}_t|X_t)$ is close to $P(Y_t|X_t)$.

\subsection{Overview of the Proposed Method} 
\begin{figure*}[!t]
\centering
\includegraphics[width=\textwidth]{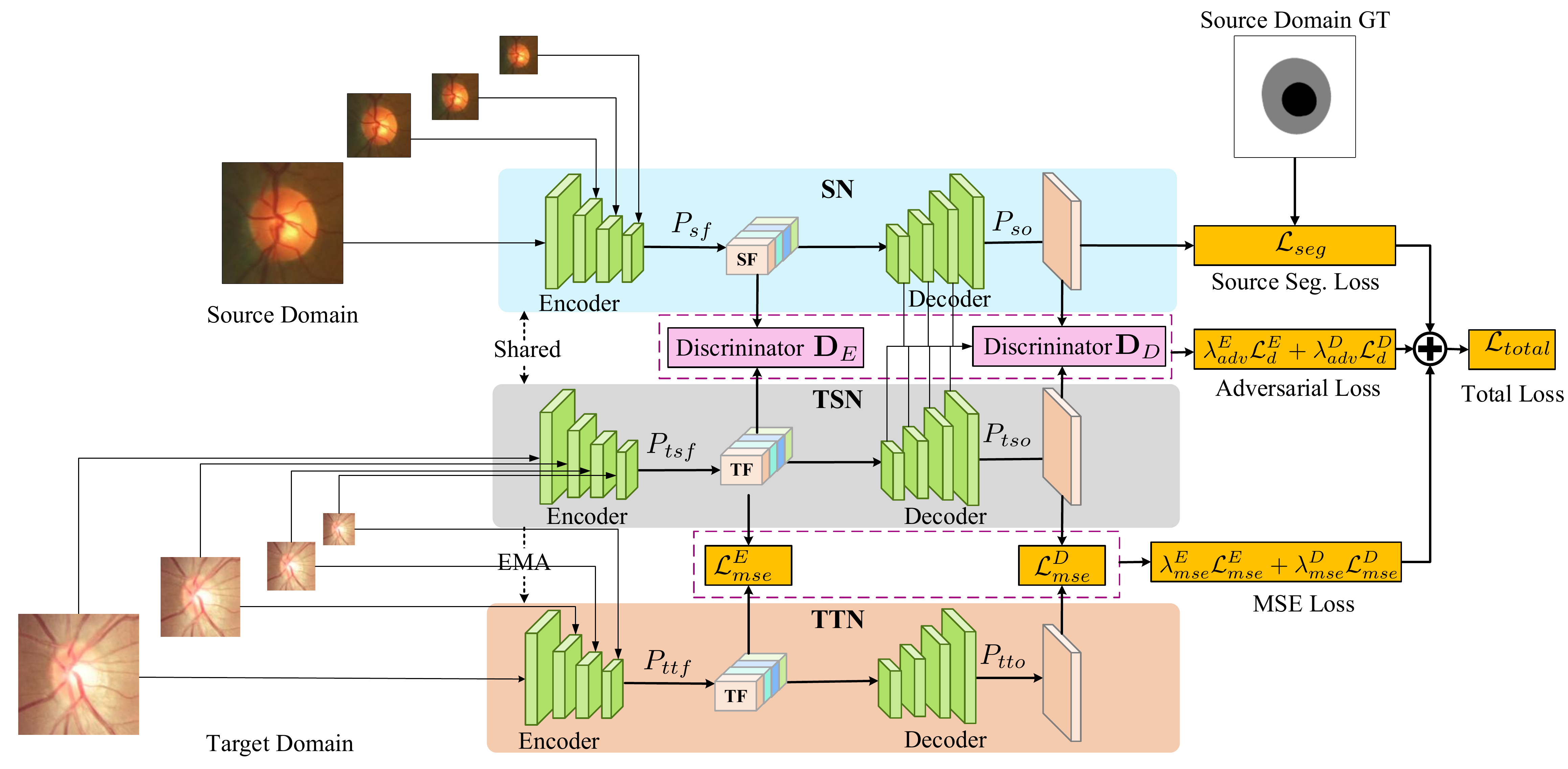}
\caption{Overview of the proposed model architecture with each primary network block colored: the Source-Domain Network (SN, in blue), the Target-domain Student Network (TSN, in gray) and the Target-domain Teacher Network (TTN, in orange). Note that discriminators can be added between all the intermediate decoder layers of SN and TSN. However, we only add the discriminators among the input ($P_{sf}$ and $P_{tsf}$) and output ($P_{so}$ and $P_{tso}$) of the decoders in this figure for simplicity.} 
\label{fig: model}
\end{figure*}

Fig.~\ref{fig: model} illustrates the design of the proposed domain adaptation framework. Our network includes three primary networks, i.e. the Source-domain Network
(SN, in blue), the Target-domain Student Network (TSN, in gray), and the Target-domain Teacher Network (TTN, in orange).  We utilize multi-scale inputs and outputs to each of the primary networks to adapt various levels of features hierarchically. During training, at each iteration, the source images are fed into SN to generate the Source-encoder Feature (SF) $P_{sf}$ and source decoder output $P_{so}$. The source domain segmentation loss is obtained by comparing the $P_{so}$ with the source domain ground truth. TSN shares the same weights with SN, and the weights of TTN are the Exponential Moving Average (EMA) weights of TSN. Adversarial losses for domain confusion are added for both encoder and decoder outputs of SN and TSN. Moreover, Mean Square Error (MSE) losses are added to both the encoder and decoder outputs of TSN and TTN.  To reduce the difficulty of high-dimensional feature calculations, the outputs of all encoders are compressed to one feature map output via a $1\times1$ convolutional layer. 

Although each of the networks plays a distinctive role in guiding networks to learn domain invariant representations, all of them can interact with each other, benefit one another, and work collaboratively as a unified framework during an end-to-end training process. SN and TSN focus on supervised learning for labeled samples from the source domain ($X_s$)
and adversarial discriminative learning for unlabeled samples from the target domain ($X_t$), separately.  More
importantly, we allow SN and TSN to share the weights that are sequentially learned from both labeled and unlabeled
samples. This technique is commonly adopted in unsupervised domain adaptation to reduce the number of learnable
parameters~\cite{tsai2018learning,zhu2019adapting}. The labeled samples enable the network to learn accurate segmentation predictions while the unlabeled samples bring unsupervised learning and further present a type of perturbation to regularize the model training~\cite{tarvainen2017mean}. 
Furthermore, TTN conducts the weight self-ensembling~\cite{french2017self} part by replicating the average weights of the TSN instead of predictions. TTN solely takes unlabeled target images as input and then the mean square difference between TSN and TTN is computed for the same target sample. Different data augmentations (e.g., adding Gaussian noise and random intensity or brightness scaling) are applied to  TSN and TTN to avoid the loss vanishing issue.

In this work, the U-Net~\cite{ronneberger2015u} with encoder-decoder structure is employed as the backbone of each network, since U-Net is one of the most successful segmentation frameworks in medical image segmentation. With the adaptability and flexibility of our framework, we expect that the results can generalize to other backbone networks and medical image analysis tasks.

\subsection{Multi-scale Input Sub-networks}
\begin{figure*}[!t]
\centering
\includegraphics[width=\textwidth]{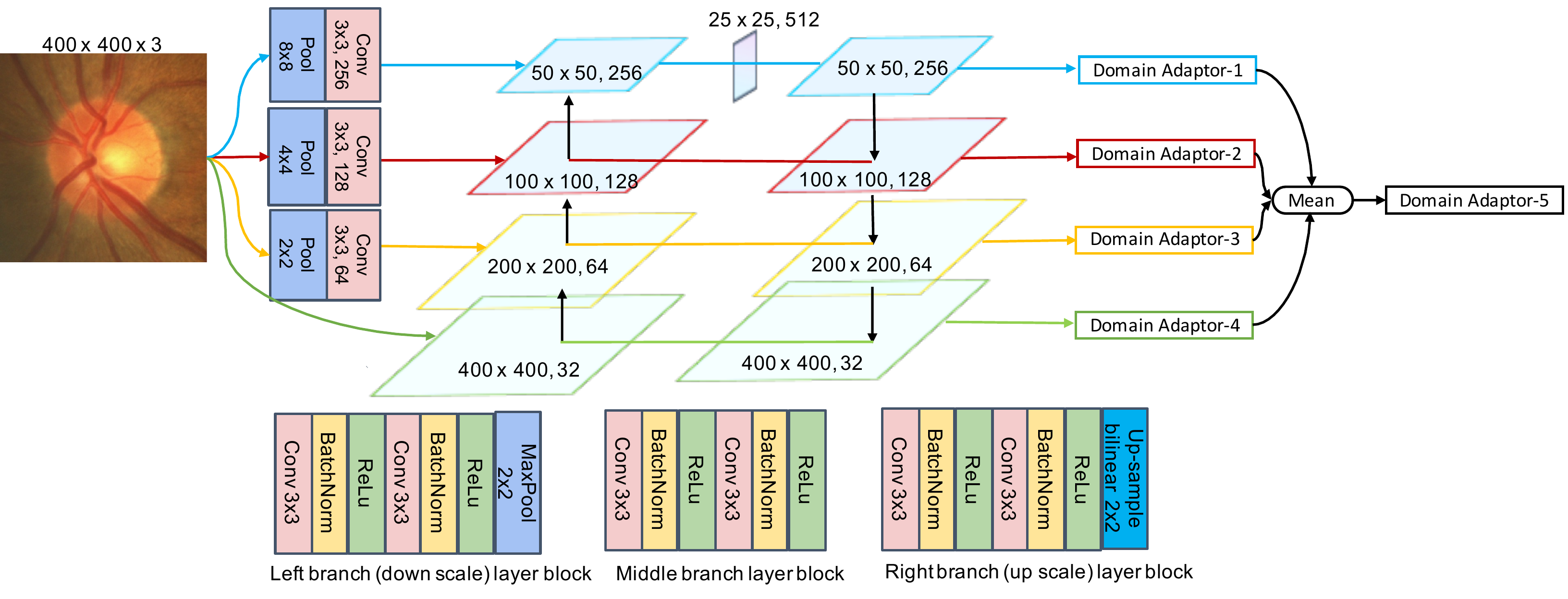}
\caption{ The network design details illustrating the multi-scale input in the combination of multiple domain adaptors applied at different hierarchical layers. Each color represents a level scale of the input that passes through several convolutional layers across both encoder and decoder of the network. The label on each layer includes the size of the input to the layer-block and the number of channels in all convolutional layers in that block.} 
\label{fig:design_multi}
\end{figure*}

We allow  multi-scale inputs to provide rich original pixel-wise features that can infuse the interaction between encoder and decoder at the different feature-learning levels of the network. Each level of the network is considered as a sub-network that primarily processes one scale of the original input for segmentation prediction. Fig.~\ref{fig:design_multi} shows the paradigm of the multi-scale input networks. To simplify the re-scaling procedure, we apply a $2\times2$, $4\times4$, and $8\times8$ size of pooling followed by a $3\times3$ convolutional layer to reduce the scale of the original input. A level scale of an input passes through several convolutional layers across both encoder and decoder of the network. Four scales of inputs are built and each of them is corresponding to a level of the neural network layers, which is considered a sub-network. The output of each sub-network uses a domain adaptor for domain adaptation. Lastly, another domain adaptor is employed as the average of the four sub-networks outputs for domain adaptation. The whole network uses multiple domain adaptors to adjust itself for domain adaption learning at both low-level and high-level features and perform an accurate segmentation prediction, especially those from the fundus optic cup and disc. There are 5 adaptors and each of them is a combination of supervised segmentation learning, the mean square difference between the prediction of the teacher and student network, and adversarial learning through deep convolutional networks as discriminators. We show the paradigm of the multi-scale input in the combination of multiple domain adaptors applied at different hierarchical layers and the details of network design in Fig.~\ref{fig:design_multi}. Notably, we include the multi-scale inputs to both the student and teacher networks of our CADA framework.

\subsection{Multiple Adversarial Discriminative Learning}
\label{sec: align_gan}

We apply five discriminators at the encoder and decoder of the networks, separately, to achieve adversarial discriminative learning. To simplify our method's explanation, the following section is using two types of discriminators for discussion: one discriminator is applied for encoder features and the other four discriminators are used for the outputs of the decoder.  The adversarial loss functions are calculated between SN and TSN. Each of the loss calculations is performed by two steps in each training iteration: (1) train the target domain segmentation network to maximize the adversarial loss $\mathcal{L}_{adv}$, thereby fooling the domain discriminator $\textbf{D}$ to maximize the probability of the source domain feature $P_{s}$ being classified as target features:
\begin{equation}
 \begin{multlined}
\mathcal{L}_{adv}(X_s) =\mathbb{E}_{x_s\sim X_s}\log(1-\textbf{D}(P_{s})),
\end{multlined}
\label{eq: adv_general}
\end{equation}
and 2) minimize the discriminator loss $\mathcal{L}_{d}$:
\begin{equation}
 \begin{multlined}
 \mathcal{L}_{d}(X_s, X_t) =\mathbb{E}_{x_t\sim X_t}\log(\textbf{D}(P_{t}))+\mathbb{E}_{x_s\sim X_s}\log(1-\textbf{D}(P_{s})),
\end{multlined}
\label{eq: dis_general}
\end{equation}
where $P_{t}$ is the target domain feature.

Note that discriminators can be added between all the intermediate decoder layers of SN and TSN. However, we only add the discriminators among the input ($P_{sf}$ and $P_{tsf}$) and output ($P_{so}$ and $P_{tso}$) of the decoders in this section for simplicity.


\subsection{Self-ensembling}
\label{sec: self_ensemble}
In self-ensembling for domain adaptation, the training of the student model is iteratively improved by the task-specific loss upon
an Exponential Moving Average (EMA) model (teacher) of the student model, which can be illustrated as:
\begin{equation}
 \begin{multlined}
    \Phi_t^{\prime}=\alpha\Phi_{t-1}^{\prime}+(1-\alpha)\Phi_t
\end{multlined}
\label{eq: ensembling}
\end{equation}
where $\Phi_t$ and $\Phi_t^{\prime}$ denote the parameters of the student network and the teacher network, respectively. EMA transfers a smooth version of the weights of the student to the teacher network. Thus, the teacher network is more stable and robust than the student. 

More specifically, at each iteration, a mini-batch of the labeled source domain and unlabeled target samples are drawn from the target domain $T$. Then, the EMA predictions and the base predictions are generated by the teacher model and the student model respectively with different augmentations applied to the target samples. Afterward, a Mean-Squared Error (MSE) loss between the EMA and target predictions is calculated. Finally, the MSE loss together with the task-specific loss on the labeled source domain data is minimized to update the parameters of the student network. Since the teacher model is an improved model at each iteration, the MSE loss helps the student model to learn from the unlabeled target domain images. Therefore, the student model and teacher model can work collaboratively to achieve robust and accurate predictions.

\subsection{Unsupervised Domain Adaptation}
\label{sec: objective}
Unlike existing methods, our method appropriately integrates  adversarial domain confusion and self-ensembling with an encoder-decoder architecture.
\subsubsection{Adversarial Feature Adaptation}
Adversarial domain confusion is applied to both the encoded features and decoded predictions between Source domain Network (SN) and Target domain Student Network (TSN) to reduce the distribution differences. According to Eq.~\ref{eq: adv_general} and~\ref{eq: dis_general}, this corresponds to the adversarial loss function $\mathcal{L}_{adv}^{E}$ for the encoder output of SN and TSN, and the adversarial loss function $\mathcal{L}_{adv}^{D}$ for the decoder output of SN and TSN:
\begin{align}
\label{eq: encoder_adv_framework}
\mathcal{L}_{adv}^E(X_s) =\mathbb{E}_{x_s\sim X_s}\log(1-\textbf{D}_E(P_{sf})),\\
\mathcal{L}_{adv}^D(X_s) =\mathbb{E}_{x_s\sim X_s}\log(1-\textbf{D}_D(P_{so})),
\label{eq: decoder_adv_framework}
\end{align}
where $P_{sf}\in \mathbb{R}^{W_e\times H_e\times C_e}$ is the encoder output and $P_{so}\in \mathbb{R}^{W_d\times H_d\times C_d}$ is the decoder output. $H_d$ and $W_d$ are the width and height of the decoders’ output; $ C_d $ refers to pixel categories of the segmentation result, which is three in our cases. $H_e$, $W_e$, and $ C_e $ are the width, height, channel of the encoders’ output. $\textbf{D}_E$ and $\textbf{D}_D$ are the discriminator networks for the encoder and decoder outputs, respectively.

The discriminator loss $\mathcal{L}_{d}^E$ for the encoder feature and the discriminator loss $\mathcal{L}_{d}^D$ for the decoder feature are as follows:
 
\begin{equation}
 \begin{multlined}
    \mathcal{L}_{d}^E(X_s, X_t) =\mathbb{E}_{x_t\sim X_t}\log(\textbf{D}_E(P_{tsf}))+\\
\mathbb{E}_{x_s\sim X_s}\log(1-\textbf{D}_E(P_{sf})) 
\end{multlined}
\label{eq: encoder_dis_framework}
\end{equation}
\begin{equation}
 \begin{multlined}
 \mathcal{L}_{d}^D(X_s, X_t) =\mathbb{E}_{x_t\sim X_t}\log(\textbf{D}_D(P_{tso}))+\\
 \mathbb{E}_{x_s\sim X_s}\log(1-\textbf{D}_D(P_{so}))
\end{multlined}
\label{eq: decoder_dis_framework}
\end{equation}
where $P_{tsf}\in \mathbb{R}^{W_e\times H_e\times C_e}$ is the encoder output and $P_{tso}\in \mathbb{R}^{W_d\times H_d\times C_d}$ is the decoder output of TSN.

\newcommand{\Req}{\textbf{Input:}\hspace*{0.5em}}
\newcommand{\X}{\hspace*{3mm}}
\newcommand{\XX}{\X\X}
\newcommand{\XXX}{\X\X\X}
\newcommand{\XXZ}{\XXX$\cdot${} }
\newcommand{\cm}[1]{$\triangleright$ #1}

\subsubsection{Collaborative Adaptation with Self-ensembling}
Self-ensembling is also applied to both the encoded features and decoded predictions between TSN and Target domain Teacher Network (TTN). In this work, the MSE is used for self-ensembling. 
The MSE loss $\mathcal{L}^{E}_{mse}$ between encoder outputs of TSN and TTN, and the MSE loss $\mathcal{L}^{D}_{mse}$ between decoder outputs of TSN and TTN can be formulated as:
\begin{align}
\label{eq: mse_framework_enc}
\mathcal{L}^{E}_{mse}(X_t) =\mathbb{E}_{x_t\sim X_t} \left [ \frac{1}{M}\sum_{i=1}^{M}(p^{tsf}_i-p^{ttf}_i)^2 \right ],\\
\mathcal{L}^{D}_{mse}(X_t) =\mathbb{E}_{x_t\sim X_t}\left [ \frac{1}{N}\sum_{i=1}^{N}(p^{tso}_i-p^{tto}_i)^2 \right ] .
\label{eq: mse_framework_dec}
\end{align}
where $p^{tsf}_i$, $p^{ttf}_i$, $p^{tso}_i$, and $p^{tto}_i$ denote the $i^{th}$ element of the flattened predictions ($P_{tsf}$, $P_{ttf}$, $P_{tso}$, and $P_{tto}$) of the student encoder, student decoder, teacher encoder, teacher decoder, respectively. $M$ and $N$ are the number of elements in the encoder feature and decoder output, respectively.

\subsubsection{Patch-based Discriminator Learning in Multi-scale Output Space}
Unlike the domain adaption for classification problems, we need to adapt both low-level and high-level features for pixel-wise image segmentation tasks. To have a better segmentation on the image across the source domain, the invariant features from both low-level and high-level layers are considered in this study. Particularly, like the study~\cite{wang2019patch}, the geometry structure of the predicted segmentation masks in the output space is used for domain adaptation. However, the dimension of the feature space from different levels of the layers varies from 256, 128, 64, to 32. Thus, we perform a 2D convolutional layer with $1\times1$ kernels on each scale of features (see Fig.~\ref{fig:design_multi}) to reduce the feature dimensions consistently to be the number of pixel classes in the segmentation mask, which is 3 classes, including background, optic disc, and optic cup. In other words, we convert the features in high-dimensional feature space to a low-dimensional output space.  We achieve domain adaptation at each level of layers of the network by applying adversarial learning on the converted low-dimensional features in the output space. We apply a patch-based discriminator on each output of the segmentation network. In the adversarial learning, the segmentation network is to fool each discriminator by producing a $m\times n\times3$ size of output having a similar distribution either from source or target domain, where $m$ and $n$ is the width and height of the output, respectively. A patch-based discriminator~\cite{isola2017image} is used to perform adversarial learning for capturing the local statistical similarity. Basically, this type of discriminator tries to classify whether each patch in a predicted mask image is following the distribution of that from the source or target domain.  More particularly, each discriminator network is composed of five convolutional layers. They have 64, 128, 256, and 512 channels, respectively. A kernel size $4\times4$ and a stride size $2\times2$ are implemented in each layer. A Leaky ReLU~\cite{xu2015empirical} layer with an alpha value of 0.2 is used after each convolutional layer. Each discriminator produces a $16 \times 16$ size output.

\subsubsection{Total Objective Function} Finally, we use cross-entropy as the segmentation loss for labeled images from the source domain. Combing Eq.~\ref{eq: encoder_adv_framework},~\ref{eq: decoder_adv_framework},~\ref{eq: encoder_dis_framework},~\ref{eq: decoder_dis_framework},~\ref{eq: mse_framework_enc}, and~\ref{eq: mse_framework_dec}, the total loss is obtained, which can be formulated as below.

\begin{equation}
 \begin{multlined}
    \mathcal{L}_{total}(X_{s}, X_{t})=\mathcal{L}_{seg}(X_s) + \lambda^{E}_{adv}\mathcal{L}_{d}^E(X_s, X_t) +  \\ \lambda^{D}_{adv}\mathcal{L}_{d}^D(X_s, X_t) +
    \lambda^{E}_{mse}\mathcal{L}^{E}_{mse}(X_t) + \lambda^{D}_{mse}\mathcal{L}^{D}_{mse}(X_t),
\end{multlined}
\label{eq: total_loss}
\end{equation}
where $\lambda^{E}_{adv}$, $\lambda^{D}_{adv}$, $\lambda^{E}_{mse}$, and $\lambda^{D}_{mse}$ balance the weights of the losses. The choice of each weight component $\lambda$ was accomplished by cross-validation in our experiments. $\mathcal{L}_{seg}(X_s)$ is the segmentation loss. Based on Eq.~\ref{eq: total_loss}, we optimize the following min-max problem:
\begin{equation}
 \begin{multlined}
    \min_{f_{\phi},f_{\tilde{{\phi}}}} \max_{\textbf{D}_E,\textbf{D}_D} \mathcal{L}_{total}(X_{s}, X_{t}),
\end{multlined}
\label{eq: min_max}
\end{equation}
where $f_{\tilde{{\phi}}}$ and $f_{\phi}$ are the source domain network with trainable weights $\tilde{{\phi}}$ and target domain network with trainable weights $\phi$. 

\section{Experiments and Results}


\subsection{Dataset}
Extensive experiments are conducted on three public datasets, including REFUGE~\cite{orlando2020refuge}, Drishti-GS~\cite{sivaswamy2015comprehensive}, and RIM-ONE-r3~\cite{fumero2011rim}, to validate the effectiveness of the proposed method. The dataset REFUGE includes 400 source domain retinal fundus images (supervised training dataset) with size $2124\times2056$,  acquired by a Zeiss Visucam 500 camera, 400 labeled (testing dataset), and 400 additional unlabeled (unsupervised training dataset) target domain retinal fundus images with size $1634\times1634$ collected by a Canon CR-2 camera. As different cameras are used, the source and target domain images have totally distinct appearances (e.g., color and texture). For other experiments, we utilize the REFUGE training dataset as our source domain for training the segmentation network and consider the Drishti-GS and RIM-ONE-r3 datasets as our target domain datasets for adaptation. The dataset Drishti-GS includes 50 images for training and 51 images for testing. The dataset RIM-ONE-r3 is split to include 99 images for training and 60 images for testing. More details of the three datasets are shown in Table~\ref{tab:dataset}. The optic disc and optical cup regions were carefully delineated by the experts. All of the methods in this section are supervised by the annotations of the source domain and evaluated by the disc and cup dice indices (DI), and the cup-to-disc ratio (CDR) on the target domain.

\begin{table}[htb]
\caption{The details of the three datasets used for the evaluation of the proposed method.}
\vspace{-15pt}
\begin{center}
\label{tab:dataset}
\resizebox{1\columnwidth}{!}{

\renewcommand{\arraystretch}{1}
\begin{tabular}{ c c c c c c }\toprule
 
Domain & Dataset & Number of images & Image size & Cameras\\
\midrule
Source  & REFUGE Train & 400 & $2124\times2056$ & Zeiss Visucam 500  \\
\midrule
\multirow{3}{*}{Target} 
 &REFUGE Train(w/o label)/Test & 400/400 & $1634\times1634$ & Canon CR-2 \\
 
 &Drishti-GS Train/Test & 50/51 & $2047\times1759$ & - \\
 
 &RIM-ONE-r3 Train/Test & 99/60 & $2144\times1424$ & - \\ 
\bottomrule
\end{tabular}
}
\end{center}
\end{table}

\begin{figure}[h]  
\centering
\includegraphics[width=\textwidth]{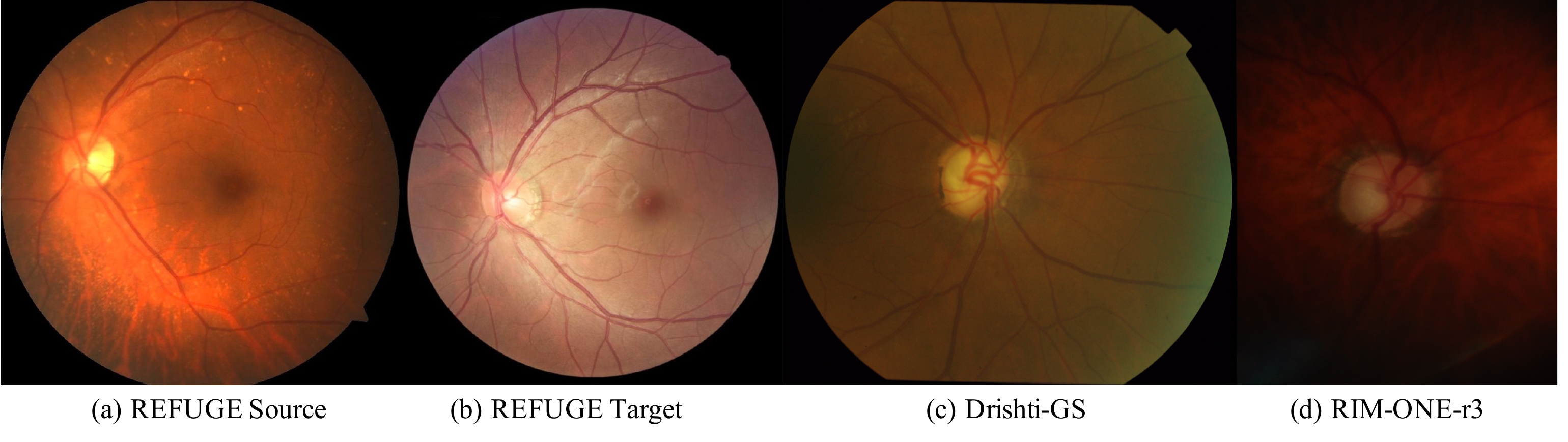}
\vspace{-20pt}
\caption{Samples from three datasets (REFUGE, Drishti-GS, and RIM-ONE-r3). The variations in color, field of vision, and textures demonstrate the challenge of domain shift. In particular, the RIM-ONE-r3 dataset carries the strongest deviation from the REFUGE and Drishti-GS datasets and thus poses a greater domain shift.} 
\label{fig: samples}
\end{figure}

\subsection{Data Preprocessing} Firstly, we detect the center of the optic disc by a pre-trained disc-aware ensemble network~\cite{fu2018joint} and then center and crop the optic disc regions. The REFUGE source domain was preprocessed to a size of $600 \times 600$ while the REFUGE target domains were pre-processed to a size of $500 \times 500$. The Drishti-GS and Rim-One datasets were preprocessed to a size of $700 \times 700$.  This is due to the different sizes of images acquired by the various cameras.

During training, all images are resized to a small size of $400\times400$ to adapt the network's receptive field. In addition, we used this pre-trained model to understand the image quality of the datasets. For example, some images' optic discs are not detectable because the image contrast is significantly low. In this case, we removed two low-quality images from the dataset RIM-ONE-r3 to evaluate our method, and the other two datasets remained as the original.

\subsection{Implementation Details} 
The U-Net is used for both student and teacher networks. Four scales of layer blocks are used for network design. Each block includes two groups of the combination with one convolutional layer, one batch normalization layer, and one ReLU layer. Each layer block is followed by either one max-pooling layer or one up-sampling layer. The details can be found in Fig.~\ref{fig:design_multi}. To train our network on the REFUGE dataset, we started from a randomly initialized U-Net and used both source and target domain images. However, for another two smaller datasets (e.g, RIM-ONE-r3 and Drishti-GS), to avoid overfitting, we first trained our network on REFUGE using the source domain images with their annotations and then applied our multi-scale domain adaptors to further train the network on the smaller datasets for domain adaptation.

We used the stochastic gradient descent (SGD) optimizer for both student and teacher networks and used the Adam~\cite{kingma2014adam} optimizer for all discriminators.  We set the initial learning rate as $1e-4$ for both student and teacher networks and adjusted the learning rate during training by $Lr_{initial}\times (1 - Iter_{current} / Iter_{max}) ^{0.9}$, where $Lr_{initial}$ means the initial learning rate, $Iter_{current}$ means the current iteration number, and $Iter_{max}$ means the maximum iteration number. We also applied polynomial decay with a power of 0.9 in a total of 200 epochs for network training. Lastly, we used morphological operations (i.e., hole filling) to post-process the results. For the discriminators, we set the initial learning rate as $2.5e-5$ and applied the same learning rate adjusting function. All experiments are processed on Python v3.6, and PyTorch v1.0.0 with NVIDIA TITAN Xp GPUs. The detailed implementations and code will be available at \url{https://github.com/cswin/CADA}.

\subsection{Evaluation Metrics}
We followed the REFUGE challenge~\cite{orlando2020refuge} and used a spatial overlap index,  dice coefficients (DI), to evaluate the segmentation performance for both OD and OC. We also used the optic cup to disc ratio (CDR) to understand the overall model performance for clinical glaucoma screening convention. The DI and CDR can be formulated as the below:
\begin{equation}
 \begin{multlined}
DI = \frac{2N^{TP}}{2N^{TP}+N^{FP}+N^{FN}}  \\ 
CDR = \frac{V^{cup}}{V^{disc}}, \quad 
\gamma_{CDR}=\left | CDR^{p}- CDR^{g}\right |
\end{multlined}
\label{eq: min_max}
\end{equation}
where $N^{TP}$, $N^{FP}$, and  $N^{FN}$ are the number of true positive, false positive, and false negative overlapped pixels, respectively; while $V^{cup}$ and $V^{disc}$ represents the vertical diameters of optic cup and disc, respectively. An absolute difference $\gamma_{CDR}$ between the predicted CDR ($CDR^{p}$) and the true CDR ($CDR^{g}$) is used for evaluation where a lower $\gamma_{CDR}$ indicates a better model performance.

\subsection{Adaptation to Datasets using Different Fundus Cameras}
Tables 2-4 lay out the segmentation results on each of the three datasets: REFUGE, Drishti-GS, and Rim-One-r3. As mentioned in the prior section, we use three metrics to evaluate the model performance, the mean dice coefficient for the optic cup (OC), the mean dice coefficient for the optic disc (OD), and the mean absolute error for the vertical cup to disc ratio ($\gamma_{CDR}$). A larger value tending to one for OD and OC means better segmentation results; for $\gamma_{CDR}$, a smaller value tending to zero represents better results. To test the effectiveness of our model, a 
``Source only'' model is trained only over the REFUGE training-source domain in a supervised manner and tested upon each respective target-testing domain.
AdaptSegNet~\cite{tsai2018learning} is one of the state-of-the-art unsupervised domain adaptation methods for image segmentation. The domain adaptation algorithm used in AdaptSegNet is also adopted similarly in the solution proposed by POSAL~\cite{wang2019patch}, the winner of the REFUGE challenge for fundus image OD and OC segmentation.  CFEA~\cite{liu2019cfea} is our previous work published in MICCAI 2019. 

\begin{table}[h]
\caption{Results of adapting the source REFUGE domain (Zeiss Visucam 500 camera) to the REFUGE target domain (Canon CR-2 camera). }
\vspace{-20pt}

\begin{center}
\label{tab: main_results}
\resizebox{1\columnwidth}{!}{
\setlength{\tabcolsep}{0.5pt} 

\renewcommand{\arraystretch}{1}
\begin{tabular}{ c c c c c c c c}\toprule
~~~~Evaluation-Index ~~&~~~~ Source only ~~~~&~~ AdaptSegNet~~\cite{tsai2018learning} ~~~~&~~CFEA~\cite{liu2019cfea}~~~~ ~~~~&~~CADA ~~~~\\
\midrule

Optic Cup & 0.7317 & 0.8198 & 0.8627 & \textbf{0.8714} \\
\midrule
Optic Disk & 0.8532 & 0.9315 & 0.9416 &  \textbf{0.9498 }\\
\midrule
$\gamma_{CDR}$ & 0.0676 & 0.0588 & 0.0481 &  \textbf{0.0447} \\
\bottomrule
\end{tabular}
}
\end{center}
\end{table} 

\begin{figure}[h]
\centering
\includegraphics[width = \textwidth]{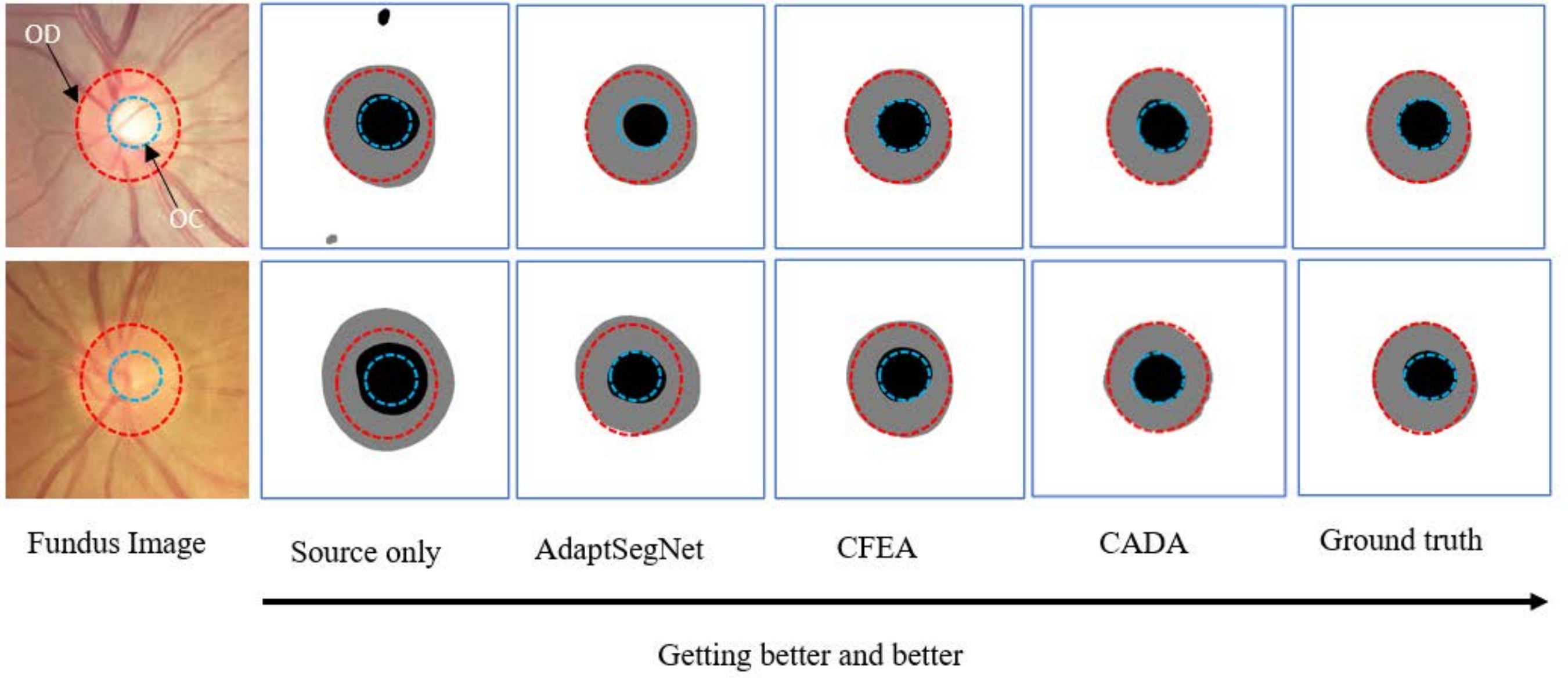}
\caption{The visual examples of the optic disc (OD) and Optic Cup (OC). The red and blue circles are the position of OD and OC in the ground truth respectively. The contours are indicative of the gap between the model results and the ground truth.}
\label{fig: mainresult}
\end{figure}

In particular, it is observed that training a network solely over the source domain is insufficient to generalize onto the new domain, thus demonstrating a key problem of domain shift. The baseline model AdaptSegNet is outperformed by CADA consistently for OD, OC, and $\gamma_{CDR}$. Furthermore, the improvement of CADA over our prior model CFEA, indicates a new paradigm for model development through multi-scale inputs and multi-discriminators. A sample visualization of the segmentation results is presented Fig.~\ref{fig: mainresult}. Overall, these results indicate that the proposed framework has the capability of overcoming domain shifts, thus allowing us to build a robust and accurate model.

\begin{table}[h]
\caption{Results of adapting the REFUGE source domain to the Drishti-GS target domain.}
\vspace{-20pt}
\begin{center}
\label{tab:Drishti_results}
\resizebox{1\columnwidth}{!}{
\setlength{\tabcolsep}{0.5pt} 

\renewcommand{\arraystretch}{1}
\begin{tabular}{ c c c c c c c c}\toprule
~~~~Evaluation-Index ~~&~~~~ Source only ~~~~&~~ AdaptSegNet~\cite{tsai2018learning} ~~~~&~~CFEA~\cite{liu2019cfea}~~~~ ~~~~&~~CADA ~~~~\\
\midrule

Optic Cup & 0.8183 & 0.8267 & 0.8271 & \textbf{0.8400} \\
\midrule
Optic Disk &  0.8800 & 0.8814 & 0.8875 &  \textbf{0.8900}\\
\midrule
$\gamma_{CDR}$ &  0.1238 & 0.1176 & 0.1133  &  \textbf{0.1110} \\
\bottomrule
\end{tabular}
}
\end{center}
\end{table}

\begin{table}[h]
\caption{Results of adapting the REFUGE source domain to the Rim-One-r3 target domain. The weaker performance relative to the domain adaptation over the REFUGE and Drishti-GS target domains is in part, considered to be due to the poor image qualities of the Rim-ONE-r3 dataset and larger domain shift.}
\vspace{-20pt}
\begin{center}
\label{tab:RIM-ONE_results}
\resizebox{1\columnwidth}{!}{
\setlength{\tabcolsep}{0.5pt} 

\renewcommand{\arraystretch}{1}
\begin{tabular}{ c c c c c c c c}\toprule
~~~~Evaluation-Index ~~&~~~~ Source only ~~~~&~~ AdaptSegNet~\cite{tsai2018learning} ~~~~&~~CFEA~\cite{liu2019cfea}~~~~ ~~~~&~~CADA ~~~~\\
\midrule

Optic Cup & 0.5916 & 0.6271 & 0.6351 & \textbf{0.6404} \\
\midrule
Optic Disk &  0.7285 & 0.7365 & 0.7506 &  \textbf{0.7664 }\\
\midrule
$\gamma_{CDR}$ &  0.1069 & 0.1017 & 0.0947  &  \textbf{0.0869} \\
\bottomrule
\end{tabular}
}
\end{center}
\end{table}

\subsection{Comparison with State-of-the-art Domain Adaptation Techniques}

We compared with AdaptSegNet~\cite{tsai2018learning}, which is one of the state-of-the-art domain adaptation methods for image segmentation. AdaptSegNet adopts adversarial learning with two discriminator networks at the last two output layers. POSAL~\cite{wang2019patch}, the winner of the REFUGE challenge, applied a similar technique into their framework for optic disc and cup segmentation at the final layer. In this study, we compared their domain adaptation technique with ours, which applies multiple domain adaptors with adversarial learning at multi-scales in both feature and output spaces. Moreover, compared to POSAL, our approach applied domain adaptation ensembles inside our framework rather than outside by averaging the results of multiple trained sub-networks. From this perspective, the primary goal of this study is to investigate a novel domain adaptation framework rather than solely pursuing higher scores by experimenting with different preprocessing techniques and backbone architectures. In particular, we find a significant improvement by extending adversarial learning not only at one scale (i.e., solely the encoder or decoder level), but rather at multiple scales, as well as our other aforementioned contributions.

To study the computational performance, we compared both CADA and POSAL on the REFUGE source and target domain experiment, the results of which are shown in Table 5. Due to the variability of factors that affect model running time~\cite{DBLP:journals/corr/CanzianiPC16}, i.e. the coding framework, structure, model depth, and GPU utilization, a holistic account is described including model procedures, model complexity, training time, and inference time. Two NVIDIA Titan Xp GPU's were utilized with a mini-batch size of 4 for 100 epochs. In particular, the CADA model was run in a PyTorch framework while the open-access POSAL model was run in a Keras-Tensorflow framework. The discriminator architectures are practically identical and as the discriminators are removed during testing, the testing time is influenced only by the network backbone. It is observed in one setting, that despite POSAL having fewer parameters than our U-NET based CADA model, our model's training and testing time is more than twice as fast. This computational discrepancy is conjectured to be influenced by the computational burden of the residual network structure~\cite{rs11151774}, the depth of the CNN's, as well as the coding framework differences (differences in parallel computing support, memory utilization, etc.). Nevertheless, the inference time suggests that our CADA model is an invaluable tool to practitioners for diagnostic purposes, being substantially faster than manual practice.

\begin{table}[h]
\caption{Computational comparison with other framework on REFUGE challenge~\cite{wang2019patch}}

\begin{center}
\label{tab: comp}

\scalebox{0.8}{

\renewcommand{\arraystretch}{1}
\begin{tabular}{ c c c} \toprule
~~~~~~&~~~~ CADA  ~~~~~&~~POSAL ~\cite{wang2019patch} ~~~~\\
\midrule 
Average Training Time Per Epoch (s) & 148.5 & 306.5 \\
\midrule
Backbone Parameters & 9.7M & 5.8M  \\
\midrule
Discriminator Parameters &  2.8M &  2.8M \\
\midrule
Average Testing Time (s) & 0.023 & 0.053 \\

\bottomrule

\end{tabular} 
}
\end{center}
\end{table}

\subsection{Ablation Study}
An ablation study is conducted solely over the REFUGE source and target domains respectively to demonstrate the influence of various components of our proposed framework.

\subsubsection{The Importance of the Encoder Adaptation (No-Enc-Ada)} 
To demonstrate the importance of encoder adaptation modules, we remove the adversarial discriminator  $D_E$ and the MSE module $mse_E$ from the encoders and then retrain the model. Fig.~\ref{fig: ablation-study} shows the performance comparison of the models with modifications on the test dataset. As one can see, without the encoder adaptation, the performance drops apparently. This comparison result may indicate that the encoder discriminative adaptation module is a crucial component for learning the domain-invariant representation.

\subsubsection{Reduce Uncertainty via Weights Self-ensembling Adaptation (No-SE-Ada)}
 
\begin{figure}[t!]
\centering
\includegraphics[width = \textwidth]{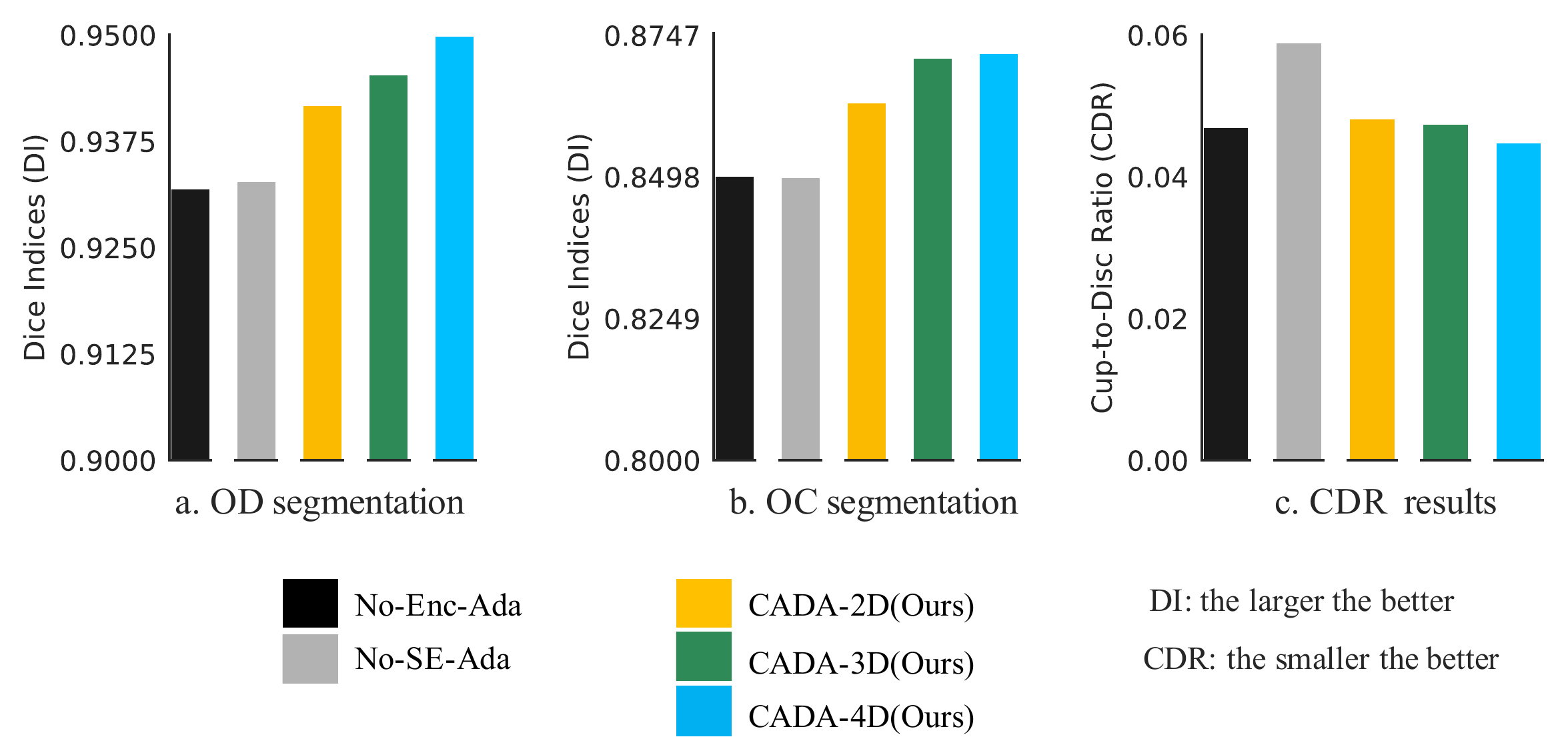}
\caption{Ablation study: performance comparison of the models with modifications on the REFUGE dataset. No-Enc-Ada means removing the discriminator from the encoder and only applying a discriminator on the decoder. No-SE-Ada means removing self-ensembling (the teacher network) from the proposed CADA.} 
\label{fig: ablation-study}
\end{figure}

We also investigate how self-ensembling adaptation affects domain adaptation performance. For this, we retrain our framework after removing the teacher network. The performance comparison of the models with modifications shows in Fig.~\ref{fig: ablation-study}. As one can see, the average performance on the test dataset is much worse than using both adversarial domain confusion and self-ensembling adaptation. Especially, for predicting the CDR, in Fig.~\ref{fig: ablation-study}-c, we can see that without weights ensembling, the CDR prediction drops down significantly. This comparison result shows that self-ensembling can significantly improve the model's robustness and the generalizability for domain shift.  More importantly, weight ensembling can reduce the model uncertainty of learning domain-invariant latent features when incorporating multiple discriminators in different feature learning space. Meanwhile, this is able to enforce all discriminators to maximize their ability to discriminate the deeper latent space features. 

\subsubsection{Multiple Discriminators Adaptation Study (CADA-2,3,4D)} 
We exploit multiple discriminators at the decoder to further investigate the maximum power of collaborative feature learning. We compare the results of applying different numbers of discriminators to different decoder layers. As one can see from CADA-2D, CADA-3D, and CADA-4D in Fig.~\ref{fig: ablation-study}, the more discriminators we use, the better result we obtain. When we apply discriminators to all decoder layers (one is at the end of the encoder, and another four are at each layer of the decoder), we obtain the best result. Notably, CADA-2D is the method proposed in our previous work  \cite{liu2019cfea}. More importantly, this comparison result further indicates that collaborative feature learning between adversarial adaptation and dynamic weight ensembling can overcome domain shift.

\begin{figure*}[t!]
\centering
\includegraphics[width = \textwidth]{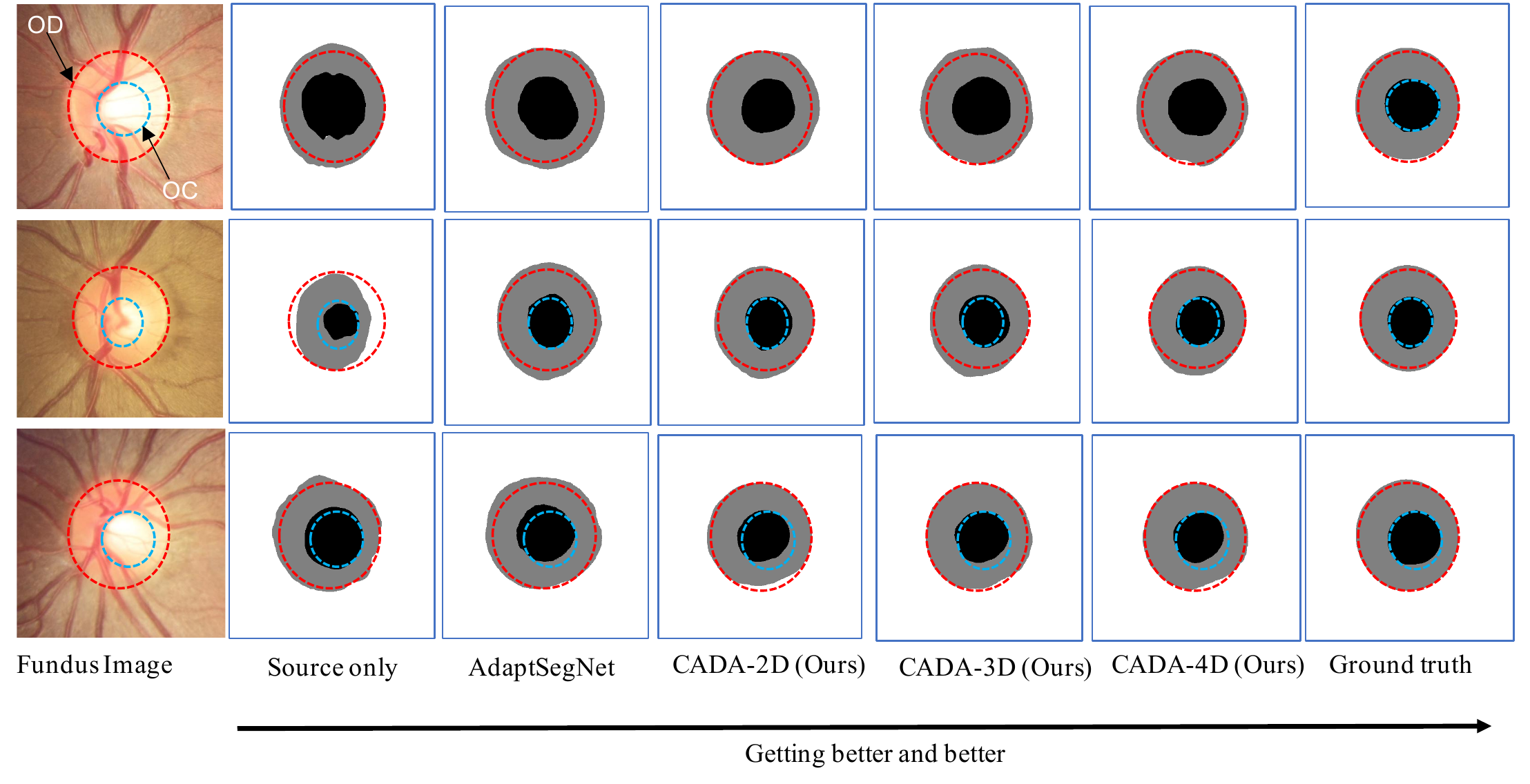}
\caption{Ablation study comparison: the qualitative examples of the Optic Disc (OD) and Optic Cup (OC) segmentation, where the black and gray region denote the cup and disc segmentation, respectively. The red and blue dash circle are the positions of the OD and OC in the ground truth, respectively. The contours are indicative of the gap between the model results and the ground truth.} 
\label{fig: ablationresult}
\end{figure*}

\subsubsection{Evaluation of $\lambda$} 
We evaluate the various combinations of $\lambda$ for balancing the segmentation, adversarial, and self-ensembling loss.  Due to the tremendous combinations, it is impossible to study all of them.  We follow the existing studies~\cite{perone2018unsupervised} and use cross-validation to investigate the most effective $\lambda$ combinations.  We find the following combination is the most effective one that can stabilize our framework training:  $\lambda_{seg} =1, \lambda^E_{adv} =0.002,  \lambda^D_{adv} =0.018,  \lambda^E_{mse} =0.057,  \lambda^D_{mse} =0.79$. 
We also show qualitative results in Fig.~\ref{fig: ablationresult} to demonstrate the effectiveness of the proposed domain adaptation model. As one can see, these qualitative results are consistent with Fig.~\ref{fig: ablation-study}. This result can further support that collaboration between adversarial learning and dynamic weight ensembling is an effective strategy to overcome domain shift in fundus images.

\section{Discussions and Conclusion}

In this work, we propose a novel method, CADA, for unsupervised domain adaptation across different retinal fundus imaging cameras, specifically over the REFUGE, Drishti-GS, and Rim-ONE-r3 datasets. 
Our CADA framework collaboratively combines multiple adversarial discriminative learning and weights self-ensembling to obtain domain-invariant features from both feature representation (encoder) and output space (decoder) in different feature scale levels. Multi-scale inputs provide hierarchical features to the collaborative learning process, while multiple domain adaptors collaboratively offer a comprehensive solution for out of distribution (OOD) samples. Weights self-ensembling stabilizes adversarial learning and prevents the network from getting stuck in a sub-optimal solution. From a complementary perspective, adversarial discriminative learning can consistently provide various model space and time-dependent weights to self-ensembling. With which, we can accelerate the learning of the domain-invariant features and conversely enhance the stabilization of adversarial discriminative learning, forming a fine collaborative circulation and generalized framework.  Moreover, we apply multiple discriminators to the multi-scale output from each layer of the decoder. These adversarial discriminative modules collaboratively encourage the encoder to extract the latent domain-invariant features. Therefore, the collaborative mutual benefits from multi-scale inputs, adversarial discriminative feature learning, weights self-ensembling, and multi-scale outputs during an end-to-end learning process, resulting in a robust and accurate model. 

Notably, the proposed framework also suggests a generalizable unsupervised learning approach. For example, we could replace the discriminator with the contrastive learning objective functions~\cite{he2020momentum,chen2020simple}. With which, the encoder can learn the rich representations rather than the invariant features. Then, we can fine-tune the encoder with limited labeled data for specific tasks, such as image classification and segmentation.  Simultaneously transferring weights with EMA from both encoder and decoder during model training is a significant novelty compared to existing representation learning methods. 

In terms of the running time, due to the multiple discriminator architecture, our framework needs relatively more computational during the training stage, compared to AdaptSegNet~\cite{tsai2018learning}, to help the segmentation network to adapt to the target domain. However, in the testing stage, the computational costs are the same as a standard U-Net network, as the images only need to go through the TTN network. Experimental results demonstrate the superiority of our domain adaptation method over other methods either by a significant performance gain or computational efficiency. Our approach potentiates a general and extendable framework to other semi-supervised and unsupervised representation learning problems.

Lastly, although we have shown marked advantages of our method, the current study has some limitations that we hope to address in the future. First, it is hard to balance the contributions (e.g., learning losses) from multiple domain adaptors during training. To find the optimal weight for each domain adaptor, we currently have to apply a massive grid-search, which can be time-consuming. Second, our framework solely investigates one type of encoder-decoder network architecture. It would be interesting to understand how our framework can improve other architectures for domain adaptation and performance across various tasks. The current work sheds light on the underlying superiority of applying multiple domain adaptors at hierarchical multi-scale feature and output space.
\section*{Acknowledgment}
Research reported in this publication is partially supported by the National Science Foundation under Grant No. IIS-1564892 and IIS-1908299, the University of Florida Informatics Institute Junior SEED Program (00129436), and the UF Clinical and Translational Science Institute, which is supported in part by the NIH National Center for Advancing Translational Sciences under award number UL1 TR001427. The content is solely the responsibility of the authors and does not necessarily represent the official views of the National Institutes of Health and the National Science Foundation.

\bibliography{refs}

\end{document}